\documentclass[prb,twocolumn,showpacs]{revtex4}
\usepackage{epsfig}

\newcommand{\startlongequation}{
\end{multicols}\vspace*{-3.5ex}{\tiny\noindent
\begin{tabular}[t]{c|} \parbox{0.493\hsize}{~} \\ \hline \end{tabular}} }
\newcommand{\stoplongequation}{
{\tiny\hspace*{\fill}
\begin{tabular}[t]{|c}\hline\parbox{0.49\hsize}{~} \\ \end{tabular}}
\vspace*{-2.5ex}\begin{multicols}{2}}
\begin{document}

\title{Strong-correlation effects in Born effective charges.}
\author{Alessio Filippetti and Nicola A. Spaldin}
\affiliation{Materials Department, University of California,
Santa Barbara, CA 93106-5050}

\begin{abstract}
Large values of Born effective charges are generally considered as reliable 
indicators of the genuine tendency of an insulator towards 
ferroelectric instability. However, these quantities can be very much
influenced by strong electron correlation and metallic behavior, which are 
not exclusive properties of ferroelectric materials. 
In this paper we compare the Born effective charges of some prototypical 
ferroelectrics with those of magnetic, non-ferroelectric compounds 
using a novel, self-interaction free methodology that improves on the 
local-density approximation description of the electronic properties. 
We show that the inclusion of strong-correlation effects
systermatically reduces the size of the Born effective charges and the
electron localization lengths. 
Furthermore we give an interpretation of the Born effective charges 
in terms of band energy structure and orbital occupations which can 
be used as a guideline to rationalize their values in the general case.

\end{abstract}
\pacs{Valid Pacs}
\maketitle


\section{Introduction}
\label{introd}

Following the implementation of the linear-response theory\cite{lrt} and the Berry phase 
approach\cite{kv,resta} to the electron polarization within the local-spin density 
approximation (LSDA), there has been tremendous progress in understanding the properties 
of piezoelectric and ferroelectric materials from a microscopic standpoint. 
Indeed, spontaneous polarization, piezoelectric constants, and Born effective charges 
(BEC), as well as full phonon spectra and potential energy surfaces can be calculated 
from first-principles giving thus a detailed picture of the electronic and structural 
behavior.

The Born effective charges (BEC), i.e. the change in electron polarization upon ionic
displacements, are the main subject of this paper. They play an important role 
in the physics of ferroelectric and even piezolectric materials, since they 
contribute to the LO-TO phonon frequency splitting at $\Gamma$ (which is 
characteristically large in ferroelectric materials\cite{zkv}) and to the piezoelectric 
tensor.\cite{bfv} Furthermore, the behavior 
of the BEC is highly non-trivial since it is related to the microscopic electron current
produced in a system by a change of the atomic position.\cite{resta_rmf} This current 
may give a contribution to the BEC which is unrelated and additional to the static 
electron charge carried by the ion during the displacement, so that the BEC can be much 
bigger than their static counterpart. In this case the BEC are usually referred to
as anomalous.
 
Interestingly, the BEC are found to be highly anomalous in the high-temperature 
(i.e. centrosymmetric) phase of many ferroelectric compounds. In particular, they 
are anomalous for the whole family of ferroelectric perovskites,\cite{gmg,rpb,prb,zkv,ggm} 
and for the class of IV-VI chalcogenides, \cite{whks}. 
Since both the BEC anomaly and the mechanism of the ferroelectric phase transition
in most of the ferroelectric perovskites can be related to the hybridization of 
anion p and cation d orbitals\cite{cok,cohen,gmg}, the interesting hypothesis arises 
that a large value of the BEC might be an indicator of the ferroelectric materials, or 
materials which are on the verge of a ferroelectric transition. 
However, according to recent calculations the BEC do not appear to posses such a 
predictive capability in general. On one hand they are found to be anomalous in a 
magnetic perovskite (CaMnO$_3$\cite{fh_camno}) which is far from ferroelectric.
On the other hand they are non-anomalous in the ferroelectromagnet YMnO$_3$, which belongs 
to the family of the hexagonal manganese oxides\cite{yakel,fiyi,aken}, 
whose ferroelectric transition is driven not by chemical activity (e.g. anion-cation 
hybridization changes) but by structural collapse and oxygen rotations.\cite{fh_science}

So we see that the meaning of the BEC and origin of the anomalous part are 
far from being completely clarified. While the role of covalent bonding in favoring 
spontaneous polarization and large BEC in ferroelectric perovskites is well 
understood,\cite{prb,cohen,gmg} other aspects which can strongly influence the 
values of the BEC, like the metallic character and the strong electron correlation 
correlation, have been scarcely investigated so far.

In an insulator on the verge of an insulating-to-metal transition, the BEC can diverge 
to anomalously large values, thus indicating an electron current which can freely 
flow through the system in response to the electric field associated with an ionic 
displacement.\cite{whks} Even more interesting is the effect of strong electron-correlation.
In a series of papers Ishihara and co-workers\cite{iet} showed that in a two-band, 
Hubbard model, the introduction of strongly-correlated effects (namely the on-site 
Coulomb energy) induces a strong enhancement of the electron-phonon coupling which becomes 
divergent at the transition boundary from the weakly-correlated (small U) to the 
strongly-correlated regime (large U). Resta and Sorella \cite{rs} showed that,
in the same model, in the region of the transition boundary, the BEC are divergent and 
undergo a sign change (i.e. the electron current changes its direction flowing from the
cation to the anion) for large U.
 
Later the same authors introduced the concept of electron localization length\cite{rs1} 
($\lambda$) as a measure to discriminate  between the metallic regime 
(where $\lambda$=$\infty$) and the insulating regime. Within the same Hubbard model, they 
found that the BEC and $\lambda$ diverge simultaneously for the same value of U 
(i.e. the BEC diverge in the metallic state),
  
Despite these predictions from model Hamiltonians, to our knowledge very few attempts 
have been made to calculate these properties for strongly-correlated materials 
from first-principles.
The main reason is likely the questionable reliabilty of LSDA calculations for
describing the electronic properties of strongly localized, atomic-like, 
electron states. It is common knowledge, indeed, that the LSDA is often 
inaccurate in describing these systems.\cite{pz,gj}
However, there is increasing evidence that many of the LSDA failures are not due 
to the lack of genuine many-body electron correlation, but rather to the presence 
of the spurious self interaction (SI) within the LSDA functional,\cite{pz} which 
becomes particularly strong for spatially localized electron charges. 
The effects of the SI were first studied by Perdew and Zunger\cite{pz} on atoms.
Subsequently Svane and co-workers\cite{svane} implemented a self-interaction-corrected 
(SIC) scheme for extended systems, obtaining large improvements over the LSDA results
for the electronic properties of transition metal oxides and high-T$_c$ superconductors.
However their method was not simple and general enough to be extensively 
applied to a vast range of systems, thus the capabilities and the performance of 
the SIC remained largely untested.

Recently we developed a self-interaction free pseudopotential scheme 
(pseudo-SIC)\cite{fh_sic} based on the idea (originally conceived by Vogel and 
co-workers\cite{vkp}) of 
expressing the SI as a non-local, pseudopotential-like projector.
This approach has the advantage of being applicable to a vast generality of systems 
including semiconductors and metals, magnetic and non magnetic materials, without
requiring large modifications or a large increase of computing effort with 
respect to the usual LSDA. The pseudo-SIC is close in spirit to the popular 
LDA+U\cite{aza,slp} since removing the SI is analogous to effectively mapping
the on-site energy U into the one-electron picture, so that both 
weakly-correlated materials and Mott-insulators can be treated on the same foot. 
In Ref.\onlinecite{fh_sic} we showed that the method is accurate in describing 
the band energies of III-V and II-VI semiconductors (e.g. GaN and ZnO), transition 
metal monoxides (NiO, MnO) and hexagonal ferroelectrics (YMnO$_3$). 

In this paper we apply the pseudo-SIC to the study of the BEC for a range of 
different materials including a prototypical ferroelectric perovskite (BaTiO$_3$), 
a ferroelectric with extremely high BEC (GeTe), a magnetic perovskites (CaMnO$_3$) 
and two magnetic monoxides (MnO and NiO). 
This investigation aims to furnish a more complete and accurate picture of the BEC, 
and to spread light on the meaning of their anomalous values.

We show that the BEC anomaly is not a property of ferroelectric materials, 
but is instead present in ionic insulators whenever the energy difference between 
filled anion and empty cation bands is particularly small (less than 3 eV), or 
in materials with nearly metallic behavior. 
Furthermore, we give evidence that the effect of strong correlation systematically 
reduces the value of the BEC and of the electron localization length. This decrease
is particularly evident for the BEC which are descibed as highly anomalous within 
LDA (or LSDA). Thus, part of (but not all) the BEC anomaly found within LDA is a 
consequence of the unphysical self-interaction which overemphasizes the orbital 
hybridization and the sensitivity of the electron-phonon coupling to the 
atomic displacement.

Finally, we propose an interpretation of the BEC in terms of occupation and capacity of 
the filled and empty orbitals, respectively, which allows a qualitative understanding 
of the anomaly in a general case.

The remainder of the paper is organized as follows: in Section \ref{form} we briefly illustrate the
pseudo-SIC scheme, and the formulation used for the BEC calculation. Section \ref{results}
is devoted to the presentation and the discussion of our results, and in
Section \ref{conclusions} we give our final remarks.

\section{Formulation}
\label{form}

\subsection{Self-interaction corrected approach}

Here we briefly summarize the main features of the pseudo-SIC approach. A detailed discussion
of the formalism and the assumptions which are at the basis of this method can be found in 
Ref. \onlinecite{fh_sic}. The formalism uses ultrasoft pseudopotentials (USPP\cite{uspp}) which
give optimal transferability and require low cut-off energies even for transition metal atoms.
The SI-free, Kohn-Sham equations are:

\begin{eqnarray}
\left[-\nabla^2 + \hat{V}_{\mbox{pp}} + \hat{V}_{\mbox{hxc}}^{\sigma}
- \hat{V}_{\mbox{sic}}^{\sigma}\right]|\psi^{\sigma}_{n{\bf k}}\rangle=
\epsilon^{\sigma}_{n{\bf k}}\,|\psi^{\sigma}_{n{\bf k}}\rangle,
\label{ks-sic}
\end{eqnarray}

where $\hat{V}_{\mbox{pp}}$ is the non-local pseudopotential,
$V_{\mbox{hxc}}^{\sigma}$ the screening potential, and $\hat{V}_{\mbox{sic}}^{\sigma}$ the SIC term.
This is expressed as a fully-non local, Kleinman-Bylander-type, pseudopotential projector:

\begin{eqnarray}
\hat{V}_{\mbox{sic}}^{\sigma}\,=\,{1\over 2}\,\sum_i\>
{|\,V_{\mbox{hxc}}^{\sigma i}\>\phi_i\,\rangle\> p_i^{\sigma}\>\langle\,\phi_i\>
V_{\mbox{hxc}}^{\sigma i}\,|\over \langle\,\phi_i\,|\,V_{\mbox{hxc}}^{\sigma i}\,|\,
\phi_i\,\rangle}\>
\nonumber
\end{eqnarray}
\begin{eqnarray}
+\,{1\over 2}\,\sum_{i,\nu,\mu}\, |\beta_{\nu}\rangle\>p_i^{\sigma}\,
\int\! d{\bf r}\>V_{\mbox{hxc}}^{\sigma i}({\bf r})\>Q_{\nu\mu}({\bf r})
\>\,\langle\beta_{\mu}|.
\label{vsic}
\end{eqnarray}

Here $i$ is a composite index which runs over angular quantum numbers and atomic positions,
$\phi_i({\bf r})$ are atomic pseudowavefunctions, $V_{\mbox{hxc}}^{\sigma i}({\bf r})$ 
are the SI potentials of the fully occupied spin-polarized atomic pseudocharges 
${\phi_i({\bf r})}^2$, and $p_i^{\sigma}$ are the corresponding occupation numbers.
 The second term in Eq. \ref{vsic} is the augmented contribution required if using USPP
($\beta_{\mu}({\bf r})$ and $Q_{\nu\mu}({\bf r})$ are projector functions 
and augmented charge desities,\cite{uspp} respectively), whereas for ordinary 
norm-conserving pseudopotentials only the first term in Eq. \ref{vsic} has to be considered.

All the ingredients needed in Eqs.\ref{vsic} are atomic quantities (thus they are 
calculated during the initialization), except the occupation numbers, $p_i^{\sigma}$, which 
must be evaluated self-consistently
by projecting the Bloch state manifold onto the pseudoatomic basis:

\begin{eqnarray}
p_i^{\sigma}=\,\sum_{n{\bf k}}\, f_{n{\bf k}}^{\sigma}\, \langle\psi_{n{\bf k}}^{\sigma}|
\phi_i\rangle \> \langle\phi_i|\psi_{n{\bf k}}^{\sigma}\rangle
\nonumber
\end{eqnarray}
\begin{eqnarray}
\times\>\left[1+
\sum_{\nu\mu}\,\langle\phi_i|\beta_{\nu}\rangle\,q_{\nu\mu}\,
\langle\beta_{\mu}|\phi_i\rangle \right].
\label{occ}
\end{eqnarray}

Again, the second term of Eq. \ref{occ} is only present for USPP. 
(The augmented charges $q_{\nu\mu}$ are integrals of $Q_{\nu\mu}({\bf r})$.)

\subsection{Electron polarization and localization}
 
For the calculation of the BEC we employ the compact formulation developed by 
Resta \cite{resta_posop}, which is based on the concept of position operator. 
This has the advantage of introducing electronic polarization and electron localization 
length as two quantities steaming from the same source. Here we describe 
the formulation for spin-polarized systems and USPP which, to our knowledge, 
has not been presented previously.

The source of the theory is the expectation value $z_{\alpha}$ of the 
many-body phase operator: 

\begin{eqnarray}
z_{\alpha}\,=\,
\left <\Psi_0\,|\,e^{i\,\Delta {\bf k}_{\alpha}\cdot\hat{{\bf R}}}\,|\,\Psi_0\right>.
\label{z_n}
\end{eqnarray}

Here $\alpha$ is the cartesian coordinate, $\hat{{\bf R}}=\sum_{i=1}^N {\bf r}_i$ 
the many-body position operator, $\Psi_0$ the ground-state wavefunction, 
and $\Delta {\bf k}_{\alpha}$ a reciprocal-space vector of magnitude:

\begin{eqnarray}
|\Delta {\bf k}_{\alpha}|\, =\, {2\pi\over \mbox{L}_{\alpha}}{1\over \mbox{M}}.
\end{eqnarray}

Here L$_{\alpha}$ and M are length and number of unit cells (that is number of grid 
points in reciprocal space) in the $\alpha$ direction, respectively.
 
Resta showed that the expected value of the position operator,
within periodic boundary conditions (PBC), must be calculated as:

\begin{eqnarray}
\left<r_{\alpha}\right> \,=\,{1\over \mbox{N} \,G_{\alpha}}\,\mbox{Im ln}\, z_{\alpha},
\label{r_a}
\end{eqnarray}

where {\it G}$_{\alpha}$=2$\pi$/L$_{\alpha}$, N is the number of electrons,
and $\mbox{Im ln}\, z_{\alpha}$ is the phase of $z_{\alpha}$. 
From Eq. \ref{r_a} the remarkable result follows that, within PBC, 
the electronic polarization P is:

\begin{eqnarray}
P_{\alpha}\,=\,{2e N\over \Omega}\left<r_{\alpha}\right>\,
=\,-{2e\over \Omega \,G_{\alpha}}\,\mbox{Im ln}\,z_{\alpha}.
\label{pol}
\end{eqnarray}

On the same footing, the electron localization, $\lambda_{\alpha}$,
(i.e. the spread of the single-particle position operator) can 
be calculated from the absolute value of $z_N^{\alpha}$ as:

\begin{eqnarray}
\lambda^2_{\alpha}\,=\,\left<r_{\alpha}^2\right>-\left<r_{\alpha}\right>^2\,=
-{1\over N\,G_{\alpha}^2}\,\mbox{ln}\,|z_N^{\alpha}|^2.
\label{loc}
\end{eqnarray}

Although Eqs.\ref{pol} and \ref{loc} are valid for any system of 
interacting electrons, $z_N^{\alpha}$ is in practice calculated within 
the usual independent-particle approximation. 
Once the many-body wavefunction, $\Psi_0$, is expressed as a single Slater 
determinant of Bloch wavefunctions we have that,\cite{note_1}

\begin{eqnarray}
z_N^{\alpha}\,=\,\mbox{det}\>\mbox{S}_{n{\bf k},m{\bf k}'}^{\uparrow}\cdot
\mbox{det}\>\mbox{S}^{\downarrow}_{n{\bf k},m{\bf k}'}
\label{z_n2}
\end{eqnarray}

where,

\begin{eqnarray}
\mbox{S}^{\uparrow}_{n{\bf k},m{\bf k}'}\,=\,
\left<\psi^{\uparrow}_{n\bf k}\,|\,e^{-i\Delta{\bf k}_{\alpha}\cdot{\bf r}}\,|\,
\psi^{\uparrow}_{m,{\bf k}'}\right>.
\end{eqnarray}
 
These are the elements of a N$^{\uparrow}\times$N$^{\uparrow}$= (N$_b^{\uparrow}$M$^3$)$\times$
(N$_b^{\uparrow}$M$^3$) matrix (N$^{\uparrow}$ and N$_b^{\uparrow}$ are the number of 
spin-up electrons and occupied bands, respectively). It is easy to see that they 
are only non-vanishing for ${\bf k}'={\bf k}+\Delta{\bf k}_{\alpha}$. Thus the determinant 
in Eq.\ref{z_n2} can be decomposed into a product of M$^3$ determinants (i.e. one for each 
point of the reciprocal space grid) of an N$_b^{\uparrow}\times$N$_b^{\uparrow}$-dimensional
matrix:

\begin{eqnarray}
\mbox{det}\>\mbox{S}_{n{\bf k},m{\bf k}'}^{\uparrow}\,=\,
\prod_{\bf k}\,\mbox{det}\,\mbox{M}_{n,m}^{\uparrow}({\bf k},{\bf k}+\Delta{\bf k}_{\alpha}),
\label{det_m}
\end{eqnarray}

where,

\begin{eqnarray}
\mbox{M}_{n,m}^{\uparrow}({\bf k},{\bf k}+\Delta{\bf k}_{\alpha})\,=
\,\left<\psi^{\uparrow}_{n,{\bf k}}\,|\,
e^{-i\Delta{\bf k}_{\alpha}\cdot{\bf r}}\,|\,
\psi^{\uparrow}_{m,{\bf k}+\Delta{\bf k}_{\alpha}}\right>.
\label{det_s}
\end{eqnarray}

A complication is caused by the use of USPP, since with the USPP method the Bloch states are not 
orthonormal, thus the expected value of the phase operator must be augmented by an 
additional term which restores the contribution of the augmented atomic charges,
q$_{ij}({\bf r})$'s. Thus, the following contribution M$_{n,m}^{US\,\uparrow}$ should be 
added to the matrix $\mbox{M}_{n,m}^{\uparrow}$ given in Eq.\ref{det_s}:

\begin{eqnarray}
{\mbox{M}_{n,m}^{US\,\uparrow}}({\bf k},{\bf k}+\Delta{\bf k}_{\alpha})\,=\,
\sum_{i,j,l}\,\left<\psi^{\uparrow}_{n,{\bf k}}\,|\,\beta_i^l\right>\,
\left<\beta_j^l\,|\,\psi^{\uparrow}_{m,{\bf k}+\Delta{\bf k}_{\alpha}}\right>\,\times
\nonumber
\end{eqnarray}
\begin{eqnarray}
e^{-i\Delta{\bf k}_{\alpha}\cdot{\bf R}_l}\,
\int\!\! d{\bf r}\>q_{ij}({\bf r})\>e^{-i\Delta{\bf k}_{\alpha}\cdot{\bf r}},
\label{aug}
\end{eqnarray}

where ${\bf R}_l$ are atomic positions.

The calculation of Eqs. \ref{det_m}, \ref{det_s}, and \ref{aug} is not demanding since the 
product over k-point space requires a dense grid only in the $\alpha$-direction, whereas 
few k-points in the subspace orthogonal to $\alpha$ are generally sufficient to obtain 
a well-converged result for electron polarization and localization length along
direction $\alpha$.

\section{Results}
\label{results}
\subsection{Technicalities}

Our local-spin density calculations\cite{pz} are performed within a plane-wave 
basis set and employ ultrasoft pseudopotentials.\cite{uspp}
For the LDA calculations, Ti (3s$^2$),(3p$^6$) semi-core 
electrons are included as valence charge in order to improve the transferability 
of their respective pseudopotentials. This is not necessary in pseudo-SIC since these 
states are much lower in energy (see Ref.\onlinecite{fh_sic} for a discussion about 
this point). In all cases we employ two beta functions for each angular channel. 
Also, non-linear core corrections have been used for all the atoms, and scalar-relativistic 
effects are included within the USPP for Ca, Ba, Ti, and Ge.

We use cut-off energies equal to 30 Ryd for CaMnO$_3$ and 40 Ryd for all the other 
compounds. For the calculation of the Berry phase, 6$\times$6 special k-point grids 
are used for the integration in the plane orthogonal to the polarization direction,
and up to 40 k-points to integrate along the parallel direction.

\subsection{BaTiO$_3$ and GeTe}

In this Section we investigate two ferroelectric materials.
BaTiO$_3$ is one of the simplest and best known ferroelectric perovskites, and the LDA successfully 
describes the energetics of the three ferroelectric phase transitions (tetragonal, orthorhombic and 
rhombohedral), provided that the experimental volumes are used for the calculations.\cite{note_batio} 
The BEC values are also well-established by a long series of LDA calculations, and
their large anomaly is assumed to be a major indication of ferroelectric instability. Also, the LDA 
calculations revealed that, despite the mainly ionic character of BaTiO$_3$, some p-d hybridization 
is present, \cite{cohen} and the sensitivity of this hybridization to off-center atomic displacements 
is a fundamental force driving the system to the ferroelectric state.\cite{cohen,gmg} 
Despite these successes, the comparison between the LDA band-structure energies and the 
photoemission spectra\cite{hudson} show typical discrepancies that are worthy of investigation.
 
\begin{figure}
\epsfxsize=9cm
\centerline{\epsffile{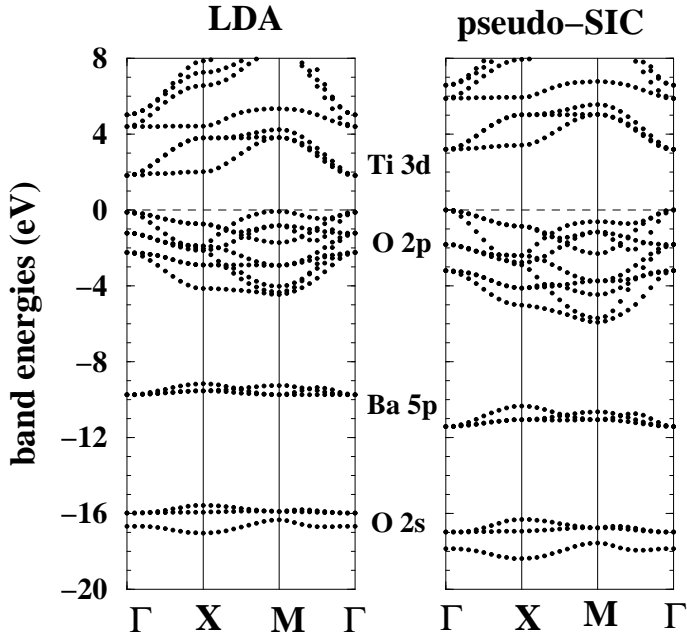}}
\label{batio_band}
\caption{Band structure of cubic BaTiO$_3$ calculated within LDA and pseudo-SIC.}
\end{figure}

In Figure \ref{batio_band} (left panel) our calculated LDA band energies of cubic BaTiO$_3$ at the
LDA volume are shown. The bands are well separated into groups, each with a single dominant orbital 
character. For the direct energy gap at $\Gamma$ we obtain the value 1.94 eV (the absolute energy gap 
is indirect between R and $\Gamma$ points and equal to 1.84 eV), in agreement with other LDA 
calculations,\cite{ggm1} while, as usual, the experiments indicate a much larger 
gap (3.2 eV\cite{wemple}). Also, the spread of the O p manifold is $\sim$ 4.5 eV, in contrast to
an experimental value of 5.5 eV.\cite{hudson} Finally, the binding energies of the more localized, 
lowest-lying states (9.9 eV for the Ba 5p bands, 16.7 eV for the O 2s bands) are
underestimated with respect to the corresponding photoemission energies (12.2 eV and 18.8 eV)

We see in the right panel of Figure \ref{batio_band} that the pseudo-SIC 
repairs the LDA deficiencies to a large extent: The energy gap at $\Gamma$ is 3.13 eV, the 
bandwidth of the O p manifold is $\sim$ 6 eV, and the energies of Ba 5p and O 2s bands 
$\sim$ 11.57 eV and 18 eV, respectively. 

There is a long-standing debate \cite{pl} about the interpretation of the mismatch between 
the LDA (or LSDA) and the observed energy gap. Briefly, if the mismatch is attributed to the 
LDA inaccuracy, this may, in principle, affect any ground-state property. If otherwise the 
mismatch steams from a 
discontinuity of the exact energy functional (i.e. it depends on the DFT itself) it has no
consequences on the ground-state properties. 
In fact, both these aspects contribute to the mismatch. In the present case, the agreement of 
pseudo-SIC and experimental energy gap is an indication that the LDA error is the 
dominant contribution.

\begin{table}
\caption{Born effective charges, Z$^*$, and localization lengths, $\lambda$, (in Bohr) 
calculated within LDA and pseudo-SIC for cubic BaTiO$_3$. 
O$_T$  and O$_P$ label oxygens on-top of Ti and planar with Ti, respectively. 
\label{tab_bec}}
\centering\begin{tabular}{cccccc}
\hline\hline
      &  $Z^*_{\mbox{Ba}}$ & $Z^*_{\mbox{Ti}}$  & $Z^*_{\mbox{O$_{\mbox{T}}$}}$  
& $Z^*_{\mbox{O$_{\mbox{P}}$}}$ & $\lambda$ \\
\hline\hline
   formal       &   2     &   4      &  --2       &   --2    &  \\
   LDA          & 2.70    &  7.01    &  --5.54    &   --2.05 & 0.93  \\
   pseudo-SIC   & 2.61    &  5.88    &  --4.43    &   --2.03 & 0.97   \\
\hline\hline
\end{tabular}
\end{table}

In Table \ref{tab_bec} the calculated values for the BEC are shown.  
The most important result is the remarkable decrease of Z$^*$ for both Ti and O$_T$,
when calculated within pseudo-SIC, with respect to their LDA counterparts. The pseudo-SIC values can 
still be described as anomalous if compared with the formal ionic charges, but the anomaly (1.88) is 
reduced by $\sim$ 40\% with respect to the LDA. This result may be expected since typically the SIC 
increases the removal energies and the space localization of the electron charges. This, in turn, 
reduces the overlap and the hybridization which is generally overemphasized within LDA.

In Table \ref{tab_bec} the calculated localization lengths, $\lambda$, are also reported.
In a many-body framework $\lambda$ is an exact measure of the electron localization. 
In the one-electron approximation, it measures the average spatial 
localization of a single electron, thus it is a useful indicator of the overall metallic 
character of a system.
The small value obtained for BaTiO$_3$ (less than 1 Bohr) is typical of highly ionic compounds 
(compare this, for example, with the values obtained for the series of III-V semiconductors in 
Ref.\onlinecite{spr}). Also, notice that the LDA value is smaller than the pseudo-SIC result. 
This is caused by the presence of Ti 3s,3p semicore states in the LDA valence charge
(these states are in the core for the SIC calculation). We will see later that, for the same 
valence states, the localization length is always smaller in pseudo-SIC. 

\begin{table}
\caption{Orbital occupation numbers for cubic BaTiO$_3$ calculated within LDA and pseudo-SIC.
The numbers show the total orbital character for individual atoms. Ba s and p orbital 
occupations are relative to the low-lying semi-core states.
\label{tab_orbocc.batio}}
\centering\begin{tabular}{cccccccc}
\hline\hline
                     & \multicolumn{3}{c} {LDA}  &  &  \multicolumn{3}{c} {pseudo-SIC} \\
\hline
                     &    Ba   &   Ti      &    O     &  &    Ba     &   Ti     &  O    \\
   s                 &   1.64  &  0.36     &   1.60   &  &   1.65    &  0.38    &  1.61   \\
   p                 &   5.63  &  1.30     &   4.82   &  &   5.65    &  1.26    &  4.99    \\
   d t$_{2g}$        &   0.81  &  1.10     &          &  &   0.72    &  0.84    &         \\
   d e$_g$           &   0.42  &  1.16     &          &  &   0.35    &  1.08    &         \\
\hline
\end{tabular}
\end{table}

The effect of the SIC on the orbital hybridizations can be appreciated in 
Table \ref{tab_orbocc.batio} where the orbital occupation numbers are reported. 
In pseudo-SIC the material is more ionic and the hybridizations 
between Ti d and O p orbitals are lowered. Indeed, in pseudo-SIC the Ti e$_g$, 
and t$_{2g}$, and the O p orbital occupations are closer to their respective 
nominal values.   

Now we present our results for GeTe.\cite{rj,raty} This is an interesting 
ferroelectric since it possesses extremely anomalous BEC\cite{whks} and shows 
electronic properties markedly different from those of the most common ferroelectric 
prerovskites. At high-temperature it has the rocksalt structure, and below 
T$_c$ $\sim$ 650 K \cite{rj} it undergoes a transition to a ferroelectric 
rhombohedral structure, with polarization parallel to the [111] axis. 

\begin{figure}
\epsfxsize=8cm
\centerline{\epsffile{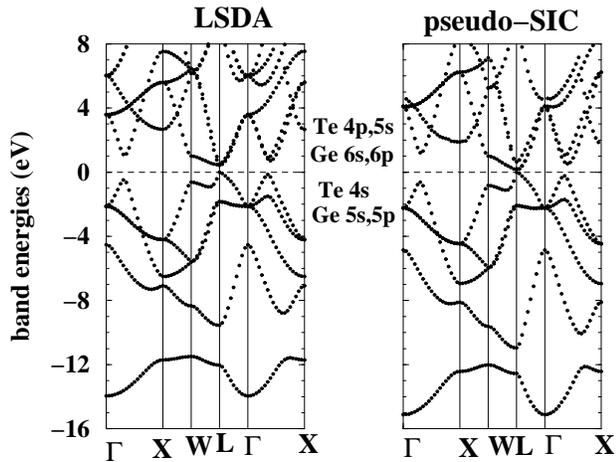}}
\label{gete_band}
\caption{Band structure of rock-salt GeTe calculated within LSDA and pseudo-SIC.}
\end{figure}

In Figure \ref{gete_band} the LSDA and pseudo-SIC band structures of rocksalt GeTe are shown.
There are no d states or other highly localized states close to the energy gap region, and in fact
the band structure shows fairly dispersed states both in the valence and in the conduction regions.
Also, in this case LDA and pseudo-SIC give very similar results. 
This is consequence of the large hybridizations of the bands around the energy gap, with s 
and p orbital characters heavily mixed. When the bands are not dominated by a well-defined 
orbital character, or are not localized on a single atomic site, the SIC corrections over 
the LDA eigenvalues are small and mainly consist of a global shift in energy of the 
whole band manifold.\cite{note_2}. 

Due to the large band dispersion, GeTe is a quasi-metal, with a direct energy 
gap at L=[1/2,1/2,1/2] 2$\pi$/a$_0$ equal to 0.43 eV in LDA and 0.20 eV in pseudo-SIC.
The metallicity is consistent with the huge LDA BEC value previously reported.\cite{whks} 
Our results give Z$_{\mbox{Te}}$= -11.55 electrons within LDA and Z$_{\mbox{Te}}$ = -8.27 
electrons in pseudo-SIC. Thus almost 35\% of the LDA anomalous charge is actually due the SI error. 
The metallic character and the effect of the SIC can be quantitatively estimated from the values 
of the localization length. In LDA we obtain $\lambda$ = 2.81 Bohr, a huge value in 
comparison to that the highly ionic BaTiO$_3$. This indicates the large (nearly metallic)
spatial extention of the valence electron charges. In pseudo-SIC we have $\lambda$ = 2.47 Bohr,
(that is 23\% smaller than the LDA value). The simultaneous decrease of $\lambda$ and BEC obtained
within pseudo-SIC is consistent with the idea that in the metallic limit, $\lambda$=$\infty$, 
the BEC calculated within the centrosymmetric structure should diverge.  

\begin{figure}
\epsfxsize=8cm
\centerline{\epsffile{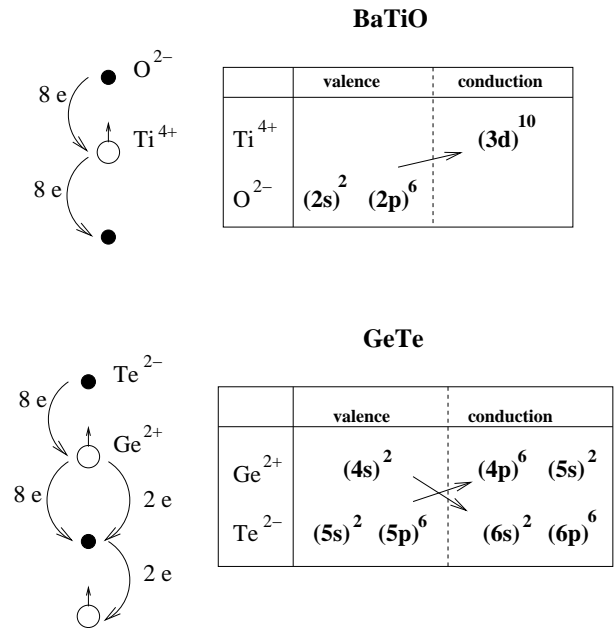}}
\caption{Schematic diagram of anomalous charge flux in BaTiO$_3$ and GeTe. 
Small arrows indicate atomic displacements, long arrows show the directions of 
electron currents. For BaTiO$_3$ the anomalous charge can only flow from filled 
O 2s,2p to empty Ti 3d orbitals and amounts to a maximum of 8 electrons. 
For GeTe two channels are available, giving a total maximum current of 10 electrons.
The calculated values are indeed below these maximum limits.}
\label{schema}
\end{figure}

To conclude this section we illustrate the mechanism of the anomalous currents in BaTiO$_3$
and GeTe by means of the simple scheme shown in Figure \ref{schema}. Consider first the upper
panel which refers to BaTiO$_3$. For simplicity, only Ti and O$_T$, which are the ions with
the most anomalous BEC, are considered. Upon the Ti displacement (indicated by the
arrow) a +4 ionic charge is transported upward. Simultaneously, a further electron current flows
from the above oxygen towards the Ti, so that the positive charge effectively carried by 
the Ti along the displacement is larger than +4. Of course, the amount of the anomalous 
contribution depends on the detail of the energy spectrum, but we can make some 
considerations based on the capacity of the orbitals available to exchange charge. 
If we assume that the current can only flow from the highest occupied O orbitals to the 
lowest unoccupied Ti orbitals\cite{note_3}, we can conclude by a simple criterion 
of orbital filling that the maximum value allowed for the anomalous contribution must be +8, 
that is (2s$^2$)(2p$^2$) oxygen electrons transferred to the 3d Ti orbitals, and than 
the largest BEC allowed according to this argument is 4+8 = +12 for Ti, and -2-8 = -10 
for O$_T$. This criterion must hold for all the octahedrally-coordinated oxide perovskites with 
a d$^0$ B-site transition metal, and indeed first-principle calculations support this 
prediction. See for example Ref.\onlinecite{gmg} where a list of BEC for many ferroelectric 
perovskites is reported. 

We can test the validity of this model for the IV-VI chalcogenides whose reported 
BEC are very anomalous,\cite{whks} the biggest of whom is indeed found in GeTe. 
In order to justify the large value of the anomalous part (9.55 in our LDA calculation) 
we should consider that there are two available channels of electron current 
(shown in the lower panel of Fig.\ref{schema}): one involves the current from the filled 
Ge (4s)$^2$ to the empty Te (6s)$^2$, (6p)$^6$ orbitals, and the other from the filled 
Te (5s)$^2$, (5p)$^6$ to the empty Ge (4p)$^6$, (5s)$^2$ orbitals.
The sum of these two contributions give 10 electrons as a maximum limit of
anomalous charge, which is indeed consistent with the calculated value.

\subsection{CaMnO$_3$}
\label{sec_camno}

In this Section we report our results for the G-type\cite{g-type} 
antiferromagnetic CaMnO$_3$.\cite{camno} 
Among the manganese perovskites, CaMnO$_3$ is the only compound stable in cubic 
symmetry since the Mn$^{4+}$ ion has the three majority d t$_{2g}$ orbitals 
fully occupied, and therefore does not undergoes Jahn-Teller distortions.
Due to its formal ionic configuration Ca$^{2+}$Mn$^{4+}$O$^{2-}_3$, this is the 
magnetic, non-ferroelectric perovskite which is better suited to be compared with BaTiO$_3$.

In a previous paper\cite{fh_camno} devoted to the analysis of orbital hybridizations 
we reported the fact that, within LSDA, CaMnO$_3$ shows anomalous BEC, with values 
similar to that of BaTiO$_3$, and we argued that the BEC anomaly could be, in fact, 
a common characteristic of the whole family of perovskite oxides. However, the 
two materials show substantial differences in the electronic properties. 
BaTiO$_3$ is definitely more ionic in character, whereas CaMnO$_3$ show a larger 
band dispersion and hybridization. Furthermore, the small band gap obtained within 
LSDA for CaMnO$_3$ suggests that the large BEC could be due to an incipient
insulator-metal transition. These facts motivated us to revisit the properties
of CaMnO$_3$ within pseudo-SIC. 

\begin{figure}
\epsfxsize=9cm
\centerline{\epsffile{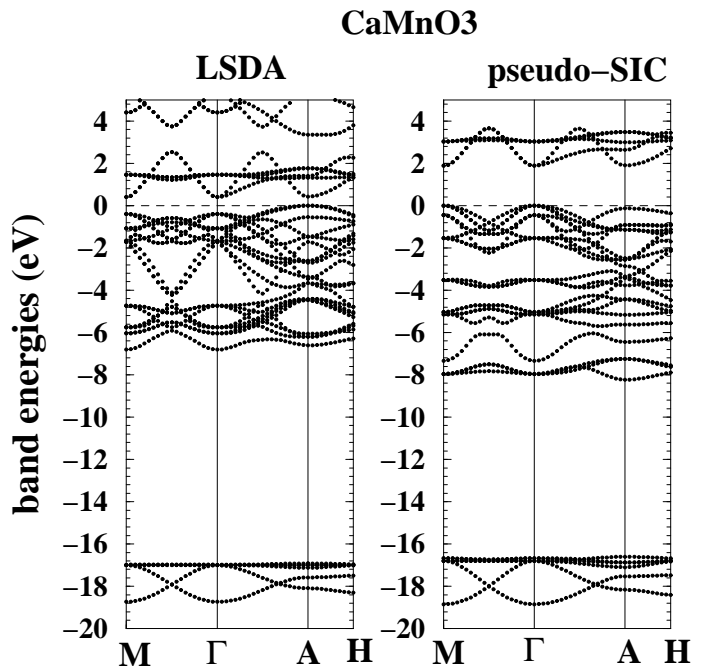}}
\caption{Band energies for G-type AFM CaMnO$_3$ calculated within LSDA and pseudo-SIC.}
\label{camno}
\end{figure}

In Figure \ref{camno} we report the band structure calculations within LSDA and
pseudo-SIC. Within pseudo-SIC the bands are flatter and spread over a wider
energy range. In pseudo-SIC the valence band manifold of O 2p and Mn d t$_{2g}$ 
states spans a region of 8 eV below the valence band top (VBT), where the same
band manifold is 7 eV wide in LSDA. In both pseudo-SIC and LSDA the O 2s state 
are placed $\sim$ 18 eV below the VBT. 
Within LSDA the fundamental energy gap is only 0.42 eV 
and occurs between points A=[1/2,1/2,0] 2$\pi/$a (the VBT) and $\Gamma$, whereas 
within pseduo-SIC the gap is direct at $\Gamma$ and equal to 1.88 eV. 
In both LSDA and pseudo-SIC the lowest group of conduction bands come from the 
majority e$_g$ and the minority t$_{2g}$ orbitals. (The t$_{2g}$ bands
are flatter and less hybridized than the e$_g$ bands since they do not point 
towards the oxigens.) 

Comparing these results with an x-ray absorption spectroscopy 
study\cite{zampieri} shows that the pseudo-SIC is in overall better agreement 
with experiment than LSDA. In particular, the pseudo-SIC reproduces well the 
observed distance between the average positions of the lowest conduction and 
highest valence bands ($\sim$ 3 eV), which is instead strongly underestimated 
in LSDA ($\sim$ 1.5 eV).
Even the observed width of the O 2p-Mn t$_{2g}$ occupied manifold ($\sim$ 9 eV)
is in better agreement with the pseudo-SIC value, whereas both LSDA and pseudo-SIC
correctly describe the material as a charge-transfer insulator (i.e. U $>$ $\Delta$
in the language of Hubbard model\cite{zampieri}).   

On the basis of the energy band results, we expect that the BEC in 
pseudo-SIC should be markedly different from the LSDA values, and the results in 
Table \ref{tab_bec_camn} confirm this expectation: the values for Z$_{\mbox{Mn}}$ and 
Z$_{{\mbox{O}}_{\mbox{T}}}$ are reduced by $\sim$ 25\% in pseudo-SIC, altough 
they still remain substantially anomalous. It is thus evident that part, but not 
all, of the BEC anomaly calculated in LSDA is due to the overestimation of 
the p-d hybridization. 
Furthermore, it is clear that anomalously large values of the BEC can
be, as in this case, absolutely unrelated to any notion of 
ferroelectric behavior. (It is helpful to remember, indeed, that the
energetics of CaMnO$_3$, disfavours any ferroelectric distortion.\cite{fh_camno})

Finally, notice that, as in BaTiO$_3$, the BEC for the A-site cation and O$_{\mbox{P}}$ 
are slightly anomalous. 

In Table \ref{tab_bec_camn} we also report the localization length. 
Within pseudo-SIC $\lambda$ is smaller by $\sim$ 20\% than in LSDA, 
consistent with the reduction obtained for the BEC.
 
\begin{table}
\caption{Born effective charges for CaMnO$_3$ calculated within LSDA and pseudo-SIC. 
For Mn we also report the individual spin-up (${Z^{el}_{Mn}}^{\uparrow}$) and spin-down 
(${Z^{el}_{Mn}}^{\downarrow}$) electronic contributions. $\lambda$ is the localization 
length (in Bohr).
\label{tab_bec_camn}}
\centering\begin{tabular}{cccccccc}
\hline\hline
  &  $Z_{\mbox{Ca}}$ & $Z_{\mbox{Mn}}$ & ${Z^{el}_{\mbox{Mn}}}^{\uparrow}$ & ${Z^{el}_{\mbox{Mn}}}^{\downarrow}$ &  $Z_{\mbox{O}_{\mbox{T}}}$  & $Z_{\mbox{O}_{\mbox{P}}}$ & $\lambda$ \\
\hline\hline
   formal        &  2      &    4    &  --3     &   0     &    --2     &    --2     &
          \\
   LSDA          & 2.74    &  7.60   &  --2.57  &  3.16   &  --6.75    &   --1.80   &
 1.42      \\
   pseudo-SIC    & 2.87    &  6.03   &  --2.73  &  1.76   &  --5.36    &   --1.77   &
 1.29     \\
\hline\hline
\end{tabular}
\end{table}

Finally, a fundamental question arises: how does the spin-polarization affect
the values of the BEC? The total (i.e. spin-integrated) BEC does not contain
the necessary information to answer this question. 
Thus, in Table \ref{tab_bec_camn} we show the electronic 
component of the Mn BEC decomposed into its spin-up and spin-down contributions 
(see Equation \ref{z_n2}), so that:

\begin{eqnarray}
Z_{\mbox{Mn}} = Z_{\mbox{Mn}}^{ion} + {Z^{el}_{\mbox{Mn}}}^{\uparrow} + 
{Z^{el}_{\mbox{Mn}}}^{\downarrow}
\label{up_down} 
\end{eqnarray}

with $Z_{\mbox{Mn}}^{ion}$ = 7. The result can be rationalized as follows: 
the majority contribution ${Z^{el}_{\mbox{Mn}}}^{\uparrow}$ is substantially 
non-anomalous and close to its formal value -3. This means that the filled 
t$_{2g}^{\uparrow}$ shell moves almost rigidly with the Mn$^{7+}$ core. 
Instead, ${Z^{el}_{\mbox{Mn}}}^{\downarrow}$ is highly anomalous.
The t$_{2g}^{\downarrow}$ states, which are at the bottom of the conduction band,
can accept at most 3 electrons of anomalous current flowing from the O onto the Mn. 
In LSDA the energy separation between the center of the t$_{2g}^{\downarrow}$ band 
manifold and the top of the O p valence bands is 
very small ($\sim$ 1 eV), thus the calculated anomalous, spin-down current (3.16) 
matches this acceptance limit (it is actually actually higher due to other secondary 
flux channels). Within pseudo-SIC, instead, the energy separation is $\sim$ 3 eV, and
the anomaly becomes drastically reduced to 1.76 electrons, that is almost half the 
LSDA value. On the other hand, the BEC contribution from the majority channel 
in pseudo-SIC is almost the same as in LSDA. 

Notice that we do not show the separated spin-up and spin-down contributions for O since 
they are equal by symmetry.

In conclusion, in a magnetic material a correct interpretation of the BEC values 
can be only achieved if spin-up and spin-down components are disentangled, since 
their contribution to the integrated value can be highly anisotropic.


\subsection{Transition-metal Monoxides}

In this section we consider two magnetic, non-ferroelectric materials which
are typical examples of systems for which the LSDA fails to give an accurate 
description, i.e. the AFM [111] A-type MnO and NiO.\cite{towk,sa,zscrhbd} 
The band structures of these 
compounds within both LSDA and pseudo-SIC have been already presented in 
Ref.\onlinecite{fh_sic}, thus they will not be shown here. 
instead, here we briefly 
summarize the main conclusions of that investigation: Within LSDA the 
fundamental energy gaps (0.9 eV for MnO and 0.4 eV for Ni) are much smaller 
than the experimental values, and are of Mott-Hubbard type (i.e. the lowest energy 
excitation occurs between filled and empty d bands). Furthermore, the local magnetic
moments (4.42 $\mu_B$ and 1.11 $\mu_B$ for Mn and Ni, respectively) underestimate
the measured values. The pseudo-SIC restores a good overall agreement
between band energies and photoemission spectra: the energy gaps are 3.98 eV for
MnO and 3.89 for NiO, the magnetic moments 4.71 $\mu_B$ for Mn and 1.77 $\mu_B$ for Ni,
and the character of the energy gap is in the intermediate charge-transfer 
Mott-Hubbard regime, in agreement with experiments.
  
In Table \ref{tab_bec_mno} the calculated values for the BEC and $\lambda$ within LSDA 
and pseudo-SIC are shown. We first discuss MnO. Perhaps surprisingly after the strong 
differences described above for the band energy structures, the BEC are found to be 
similar and substantially non anomalous within LSDA and pseudo-SIC (our LSDA result 
is in perfect agreement with an earlier linear-augmented plane waves calculation \cite{mpbr}).
This result can be understood from the band structures shown in 
Ref.\onlinecite{fh_sic}: MnO has, nominally, five fully occupied majority d bands and 
five empty minority d bands. In LSDA these two manifolds are separated by Hund's 
rule, whereas in pseudo-SIC the on-site Coulomb energy is dominant and the 
d-d energy gap increases dramatically. However, for this material the parameter 
which controls the anomalous charge is not the fundamental gap, but the energy 
separation between the highest occupied O p bands and the lowest unoccupied 
Mn d band. This is large ($\sim$ 4 eV) and similar in both LSDA and pseudo-SIC, 
thus the non-anomalous value for the BEC are justified.

Now consider the BEC decomposition within ionic and electronic, up-spin and 
down-spin components (Equation \ref{up_down}). The up-spin and down-spin 
formal charges are -5 and 0.  We see that in LSDA both up-spin and down-spin 
electronic contributions are slightly anomalous, but the anomalies in the two 
channels have opposite signs (that is the currents flow in opposite directions) 
thus they cancel each other in the integrated value. The pseudo-SIC, in fact, 
reduces the anomalies in both channels, but, due to the current cancellations, 
this is not revealed in the total BEC which is indeed very close to the LSDA value. 

\begin{table}
\caption{Born effective charges for MnO and NiO calculated within LSDA and pseudo-SIC. 
The individual spin-up and spin-down electronic contributions are also shown. 
$\lambda$ is the localization length (in Bohr).
\label{tab_bec_mno}}
\centering\begin{tabular}{ccccc}
\hline\hline
 MnO & $Z_{\mbox{Mn}}$ & ${Z^{el}_{\mbox{Mn}}}^{\uparrow}$ & ${Z^{el}_{\mbox{Mn}}}^{\downarrow}$  & $\lambda$ \\
\hline
   formal        &   2      & --5      &   0     &                   \\
   LSDA          &  2.31    & --5.13   &  0.44   &  1.36        \\
   pseudo-SIC    &  2.29    & --4.96   &  0.25   &  1.19        \\
\hline
    NiO &   $Z_{\mbox{Ni}}$ & ${Z^{el}_{\mbox{Ni}}}^{\uparrow}$ & ${Z^{el}_{\mbox{Ni}}}^{\downarrow}$ & $\lambda$ \\ 
\hline
   formal        &   2      & --5      &  --3     &          \\
   LSDA          &  2.31    & --4.47   &  --3.22  &  1.18     \\
   pseudo-SIC    &  2.04    & --4.87   &  --3.09  &  0.76     \\
\hline\hline
\end{tabular}
\end{table}

For NiO we can repeat most of the considerations expressed for MnO. 
Now the energy gap in LSDA lies in the minority d band manifold 
and is due to the crystal field splitting which is only $\sim$ 1 eV. 
This value underestimates by far the experimental value. However, as in MnO
the p-d energy separation is $\sim$ 4 eV and therefore no large anomalous charges 
contribute to the BEC. From the up-spin and down-spin decomposition 
(now $Z_{\mbox{Ni}}^{ion}$ = 10) we see that there is also a cancellation 
between the corresponding anomalies. The cancellation is almost complete 
within pseudo-SIC, thus the BEC is almost equal to the formal value.  

Finally, within pseudo-SIC we obtain a significant decrease of localization lengths 
for both these compounds with respect to the LSDA.

\section{Discussion}
\label{discussion}

Finally, we summarize the main findings of our analysis.
We have seen that including strong-correlation effects within the band energy 
structure of strongly-correlated materials sensitively affects the values of the BEC 
and the electron localization lengths. With respect to the LDA or LSDA results, 
the pseudo-SIC systematically reduces both BEC and $\lambda$ for all the investigated
materials. This effect is particularly evident for materials with highly anomalous BEC 
(i.e. BaTiO$_3$ and CaMnO$_3$), and can be understood as resulting from the fact 
that the LSDA overemphasizes the orbital hybridizations and therefore the electron hopping.
The decrease of electron localization lengths is also expected since the SIC 
enhances the binding energies and therefore the spatial localization of the electron charges. 
However, we point out that even within pseudo-SIC the anomalous contribution to the 
BEC is still substantial.

Our investigation reveals that the BEC are controlled by two fundamental parameters: 
one is the the energy separation between the highest occupied anionic bands and the 
lowest empty cationic bands. This criterion explains the trend of the BEC calculated 
for BaTiO$_3$ and CaMnO$_3$ (where the separation is less than 2 eV in LSDA and 
$\sim$ 3 eV in pseudo-SIC) and the transition-metal mono-oxides, where it is $\sim$ 4 eV 
in both LSDA and pseudo-SIC. The other important element is the metallic character of the
system. It is very sound that in the metallic limit the BEC must be divergent,
and the electron localization length gives us the possibility of evaluating this limit 
quantitatively. GeTe is a perfect material for such an analysis, since it is a ferroelectric
with highly anomalous BEC and nearly metallic behavior. We found the GeTe energy gap small 
in both LDA and pseudo-SIC, but in pseudo-SIC both the localization length and the BEC 
are markedly smaller.

Finally, we should comment on the results of the two-band Hubbard model 
calculations,\cite{iet,rs} since our results are in contrast with
these works, which predict that the introduction of the on-site Coulomb 
energy, U, should increase the BEC with respect to a single-particle scheme, 
and the BEC should undergo a sign change corresponding to the 
charge-transfer to Mott-Hubbard transition. 

The disagreement is due to the fact that in the Hubbard model it is
assumed that the on-site Coulomb energy of the anion favors a ground state with 
mixed anion-cation occupancy, instead of a purely anionic ground state.
This causes, at large U, the transition from the charge-transfer to the
Mott-Hubbard regime, and also produces a sign change in the BEC since now 
the anomalous current can flow from the filled cation to the empty anion.

In density fuctional calculations what happens is quite the opposite: 
with the on-site Coulomb interaction on the ionic site turned on, a 
larger energy cost is associated with the fluctuation of the respective occupancy. 
The tendency to fluctuation is reduced by lowering (not raising) the energy of the filled 
anion bands. Thus, strong correlation effects further favor charge-transfer 
ground states over Mott-Hubbard regimes. This is in complete agreement with the 
photoemission spectroscopy data\cite{towk,zscrhbd} for transition-metal and 
perovskite oxides, as well as with the LDA+U results.\cite{aza,st,babh} 
In conclusion, we think that the two-band Hubbard model is not adequate to 
describe materials like the ones studied in this paper, which present a 
complicated cohexistence of charge localization and hybridization.


\section{Conclusions}
\label{conclusions}

In this paper we revisited the behavior of the BEC by means of our recently proposed 
pseudo-SIC method which set the LSDA density functional free of the spurious SI, 
and thus describes the electronic properties of weakly as well as strongly correlated 
materials efficiently. We investigated two ferroelectric materials with highly anomalous 
BEC, and some magnetic materials with both anomalous and non-anomalous BEC.

Overall, we find that the anomalous part is systematically reduced 
(but not suppressed) by the SIC. The electron localization lengths is 
also reduced by the SIC, consistently with the idea that in the metallic 
limit the BEC must be divergent.

On the basis of these considerations, what can be argued about the BEC's ability
to predict a ferroelectric instability? According to our results, there is no evidence 
to claim such a role for the BEC, since their values are related to features of the 
band energy spectrum which are not specifically properties of a ferroelectric material. 
We may argue that a compound with large BEC, if ferroeoectric, will have a large 
spontaneous polarization, since this represents the integrated value of the BEC over 
the ferroelectric displacement,\cite{resta_rmf} but the BEC value 
in itself is not indicative of, or specific to a chemical environment particularly 
favorable for non-ferroelectric-to-ferroelectric phase transition.

\acknowledgments

We thank Umesh Waghmare for many thoughtful discussions.
We acknowledge financial support from the National Science Foundation's Division of Materials
Research under grant number DMR 00-80034. Most of the calculations have been carried out on 
the IBM SP2 machine of the MHPCC Supercomputing Center in Maui, HI.




\begin{thebibliography}{99}
\bibitem{note_1}
We use the fact that the overlap between two single Slater determinants equals 
the determinant of the matrix whose elements are overlaps of occupied Bloch states.
Also, we assume that the Slater determinant do not couple spinors with different 
directions (thus, no spin-orbit interaction is present), so that up and down 
determinants can be factorized.
\bibitem{note_batio}
As usual, the volume calculated within LDA is too small. If the LDA volume is
assumed, BaTiO$_3$ is not ferroelectric.
\bibitem{note_2}
We point out that the pseudo-SIC does not account for genuine many-body corrections.
For example in bulk silicon the band energies in pseudo-SIC are substantially identical 
to the bands calculated within LDA.
\bibitem{note_3}
Indeed, this is a simplification since other orbitals such as Ba 5s 
and 5p furnish small contributions as well. See the band-by-band 
resolution for the BEC of BaTiO$_3$ calculated in Ref.\onlinecite{gmg}.
\bibitem{g-type}
In the G-type configuration, all nearest-neighbor Mn have antiparallel
spin allignement.
\end{thebibliography}
\end{document}